\documentclass[prl,amsmath,amssymb,dvips,showpacs]{revtex4}
\usepackage{graphics,epsfig,amssymb}

\def\beq{\begin{equation}} 
\def\eeq{\end{equation}} 
\begin{document}

\title{Isoscalar-isovector proton-neutron pairing and  quartet condensation in $N=Z$ nuclei}

\author{M. Sambataro$^a$ and N. Sandulescu$^b$}
\affiliation{$^a$Istituto Nazionale di Fisica Nucleare - Sezione di Catania,
Via S. Sofia 64, I-95123 Catania, Italy \\
$^b$National Institute of Physics and Nuclear Engineering, P.O. Box MG-6, 
Magurele, Bucharest, Romania}

\begin{abstract}
 We show that the correlations generated in the ground state of $N=Z$ nuclei by the isovector and  isoscalar  pairing
 forces can be treated with high precision as a condensate of alpha-like quartets. To treat these correlations, 
the quartet condensation model (QCM) is extended to the treatment of spherically symmetric 
isovector $(T=1,J=0)$ and isoscalar $(T=0,J=1)$ pairing forces .  Within QCM we discuss 
the competition between $T=1$ and $T=0$ pairing correlations in the case of a two-level model and for $N=Z$ nuclei 
with  nucleons moving in the open shells above $^{16}$O, $^{40}$Ca and $^{100}$Sn. We show that in $N=Z$ systems
isovector and isoscalar proton-neutron pairing correlations always coexist.
\end{abstract}

\maketitle

\section{Introduction}
 
 One of the long standing questions in nuclear structure is whether a condensate of
 deuteron-like proton-neutron pairs could exist in $N=Z$ nuclei, in analogy with the condensates of like-particle pairs which account
 rather well for the  pairing correlations in heavy $N>Z$ nuclei. For systems in which
 the pairing force acts only among one type of nucleons, the ansatz of the pair condensate, on which all BCS-type models are based,  is supported by the fact that the pair condensate is the exact ground state of the pairing Hamiltonian
 for  nucleons moving in degenerate orbits. When the levels are not degenerate
 and the pairing force is state-dependent, the exact solution is not anymore a condensate of identical pairs
 but rather a product of distinct (complex-conjugate) pairs \cite{richardson} or, equivalently, a product of distinct (real) quartets built
 by four identical particles \cite{sasa_su2}. However, as shown already many years ago \cite{bayman,dietrich}, the ansatz
 of the pair condensate  (usually known as  projected-BCS approach (PBCS)) provides a good approximation  
even for non-degenerate levels and general state-dependent  interactions, under the condition that nucleons move in orbits close to the Fermi level \cite{sandulescu_bertsch}. 
 
  Based on the success of BCS-type models for like-particle pairing, these models has been applied,
   in a generalized form, to the treatment of pairing between neutrons and protons in $N=Z$ nuclei  (e.g., see
   the review paper \cite{goodman} and the references quoted therein). However, unlike the case of like-particle pairing, 
  the ground state of protons  and neutrons moving in degenerate levels  cannot be described exactly as a pair 
  condensate. Indeed, as shown by the SU(4) model for isovector-isoscalar pairing , in the case of degenerate levels the 
  ground  state of  $N=Z$  systems is exactly described by a condensate of alpha-like quartets rather than by a 
  condensate of pairs \cite{dobes}. 
   Quartet-type correlations have been also identified numerically in
  the exactly solvable SU(4) model for non-degenerate levels of angular momentum $l=0$ \cite{lerma_so8}. 
  The SU(4) pairing model is, however, not realistic enough because it neglects the role played by the spin-orbit  force, 
  which is crucial for the description of atomic nuclei.  Various studies have shown that the spin-orbit force strongly hinders the proton-neutron pairing correlations in the ground and excited states of $N=Z$ nuclei \cite{pinedo,lei_pittel,bertsch_luo}. 
  Yet, the  majority of these studies do not discuss  the types of correlations induced by the proton-neutron
  pairing in the presence of spin-orbit interaction.  In Refs. \cite{qcm_t1,qcm_ngz_t1,qcm_wigner} 
  it was shown that the ground state of realistic isovector pairing
  Hamiltonians can be described with high precision as a condensate of  alpha-like 
   quartets, as suggested by the exactly solvable SO(5) model \cite{dobes}. However, at variance with the SO(5) model, the quartet condensation 
   model (QCM) for realistic pairing Hamiltonians is based on  collective quartets which take properly into account the 
   spin-orbit interaction. 
          
     The  isovector pairing can also be  described within a quartet model in which the ground state is represented
     not as a product of identical quartets, as in QCM, but as a product of distinct quartets \cite{qm_t1}.
      This quartet model (QM),  which proposes a ground state which is  analogous to that of Richardson for like-particle pairing,  has been recently generalized to treat
      the isovector-isoscalar pairing interaction \cite{qm_t1t0} and has also been employed to analyze four-body 
      correlations in  nuclei \cite{qm_prl}.    
  
    The scope of the present paper is to extend the QCM formalism to the treatment of the isovector-isoscalar pairing interactions, to analyze 
    the accuracy of this approximation with respect to the exact solution of realistic pairing Hamiltonians as well as to the QM approach
    and to discuss, within QCM, the effect of spin-orbit  interaction on the competition between the isovector and isoscalar paring correlations. 
    In this study we will focus on spherically symmetric pairing interactions.  An extension of the QCM  approach for the axially deformed 
    isovector-isoscalar pairing interactions  was  recently proposed in Ref. \cite{qcm_def}.  
    
   \section{Formalism}

 The isovector and isoscalar  pairing correlations in nuclei are usually studied with the Hamiltonian
\begin{equation}
H=\sum_i  \epsilon_i N_i+ \hat{V}_{so}  +  g_1 \sum_{i, k,T_z} \mathcal{P}^+_{i,T_z} \mathcal{P}_{k,T_z}+
g_0 \sum_{i, j,S_z}  \mathcal{D}^+_{i,S_z} \mathcal{D}_{k,S_z}.
\end{equation}
In the first term $\epsilon_i$ and $N_i$ are, respectively, the energy and the particle number operator relative to the single-particle 
state $i=\{n_i,l_i,\tau_i\}$, where $l_i$ is the orbital angular momentum and $\tau_i$ is the isospin projection.
The Coulomb interaction between the protons is not taken into account so that the single-particle energies of  protons  and neutrons
are assumed to be equal. The second term in Eq. (1) is the  spin-orbit interaction for protons and neutrons, which has the standard
expression. The third and the fourth terms are, respectively, the   isovector $(T=1, S=0)$ and isoscalar $(T=0,S=1)$ pairing interactions. 
They are written in terms of  the pair operators
\begin{equation}
\mathcal{P}^+_{i,T_z}= \sqrt{\frac{2l_i+1}{2}}[a^+_i a^+_i ]^{T=1,S=0,L=0}_{T_z}
\end{equation}
\begin{equation}
\mathcal{D}^+_{i,S_z}= \sqrt{\frac{2l_i+1}{2}}[a^+_i a^+_i ]^{S=1,T=0,L=0}_{S_z} ,
\end{equation} 
where $L$,  $S$ and $T$ are the orbital momentum, the spin and the isospin of the pairs, respectively.
When the spin-orbit is neglected and the orbits are degenerate, the Hamiltonian (1) has the SO(8) symmetry. If, in addition,  $g_1=g_0$, the Hamiltonian (1)
has the SU(4) symmetry and can be solved analytically both for degenerate and non-degenerate levels \cite{dobes,lerma_so8}. This is not anymore possible
in the presence of the spin-orbit interaction.  

The question we address in this study is whether the ground state of the  Hamiltonian (1) as well as of the most general
isovector-isoscalar pairing  Hamiltonian (17) (see below), can be well approximated by a condensate of alpha-like quartets, 
as in the case of isovector pairing \cite{qcm_t1}. Thus, as in Ref.\cite{qcm_t1},  we represent the ground state as a product of identical quartets
\begin{equation}
|\Psi_{gs}\rangle  = (Q^+)^{n_q} |0 \rangle. 
\end{equation}
The quartet operator $Q^+$ is taken as a sum of two  quartets
\beq 
Q^+=Q^+_1+Q^+_0,
\eeq
where  $Q^+_1$ is the collective isovector quartet formed by coupling two isovector pairs to total $T=0$, i.e.,  
\beq
Q^+_1= \sum_{j_1 j_2} x_{j_1 j_2} [P^+_{j_1}P^+_{j_2}]^{T=0}
\eeq
and $Q^+_0$ is the collective isoscalar quartet built by coupling two isoscalar pairs to total $J=0$, i.e.,
\beq
Q^+_0= \sum_{j_1 j_2 j_3 j_4} y_{j_1 j_2 j_3 j_4} [D^+_{j_1 j_2}D^+_{j_3 j_4}]^{J=0}.
\eeq
 These quartet operators are expressed in terms of the pair operators in $jj$ coupling scheme:
 \begin{equation}
P^+_{j,T_z}= \frac{1}{\sqrt{2}}[a^+_j a^+_j ]^{T=1,J=0}_{T_z}
\end{equation}
\begin{equation}
D^+_{j_1 j_2 J_z}= \frac{1}{\sqrt{1+\delta_{j_1j_2}}}[a^+_{j_1} a^+_{j_2} ]^{J=1,T=0}_{J_z}.
\end{equation} 

In Ref. \cite{qcm_t1}, the QCM state was further simplified by factorizing the mixing amplitudes which define the quartets. Due to this factorization
it was possible to express the quartet condensate in terms of collective pairs and  to use the recurrence relations method for  the evaluation of the 
expectation value of the Hamiltonian. If one adopts the same factorization in the present formalism, therefore assuming that  
$x_{j_1 j_2}=\bar{x}_{j_1} \bar{x}_{j_2}$ and 
$y_{j_1 j_2 j_3 j_4}=\bar{y}_{j_1 j_2} \bar{y}_{j_3 j_4}$, the collective quartets can be written as
\beq
\bar{Q}^+_1 =2\Gamma^+_1 \Gamma^+_{-1}- (\Gamma^+_0)^2
\eeq
\beq
\bar{Q}^+_0= 2\Delta^+_1 \Delta^+_{-1} -{\Delta^+_0}^2 .
\eeq
These quartets are expressed in terms of  the  collective isoscalar and isovector pairs 
\beq
\Gamma^+_{T_z} = \sum_j \bar{x}_j P^+_{j, T_z}
\eeq
\beq
\Delta^+_{J_z} =\sum_{j_1 j_2} \bar{y}_{j_1 j_2} D^+_{j_1 j_2 J_z} .
\eeq
It is soon realized that, when formulated in terms of collective pairs, the wave function (4) becomes a complicated superposition of
mixed condensates, formed by all type of pairs. If the isoscalar quartet is further reduced to include only the $\Delta^+_0$ pairs, this formalism
becomes formally equivalent to the one proposed in Ref.\cite{qcm_def} for the treatement of the isovector-isoscalar pairing forces acting on axially 
deformed states.
 
The collective isovector and isoscalar pairs defined above can be  used to construct various 
PBCS-type states for $N=Z$ systems. Thus, with the isovector pairs (12) can be formed the following PBCS states with well-defined 
numbers of protons and neutrons \cite{qcm_t1}:
\beq
| PBCS1 \rangle = (\Gamma^+_1 \Gamma^+_{-1})^{n_q} | 0 \rangle 
\eeq
\beq
| PBCS0_{iv} \rangle = (\Gamma^+_0)^{2n_q} | 0 \rangle.
\eeq
Both states have, as required, $J=0$ and $T_z=0$, but they do not have a well-defined total isospin. 
Similar PBCS states can be constructed with the isoscalar pairs (13). Of physical interest is the PBCS state
\beq
| PBCS0_{is} \rangle = (\Delta^+_0)^{2n_q} | 0 \rangle.
\eeq
This state has $T=0$ and  $J_z=0$, but it has not a well-defined angular momentum. 
Since the states (15) and (16) are condensates, respectively, of $T=1$ and $T=0$ proton-neutron pairs, one might think that a comparison 
of their correlation energies could give a clear evidence on what type  of proton-neutron pairing is prevailing in $N=Z$ nuclei.
However, a conclusion based only on this
comparison would be misleading because, as shown in the next section, the PBCS approximation  is not accurate 
enough to describe properly the isovector and isoscalar pairing correlations.

In this work we will consider  the case in which the mixing amplitudes $x_{ii'}$ and $y_{ii'jj'}$ are factorized, as discussed above,
and also the case in which they keep their original form. In both cases these amplitudes will be constructed variationally by minimizing the 
expectation value of the pairing Hamiltonian in the QCM or PBCS-type states.

The QCM formalism proposed in this paper can also be  applied to the most general spherically symmetric 
isovector $(T=1,J=0)$ and isoscalar $(T=0,J=1)$ pairing forces described by the Hamiltonian
\begin{equation}
H=\sum_i  \epsilon_i N_i + 
\sum_{i,j} V^{T=1}_{J=0} (i,j) \sum_{T_z}P^+_{i,T_z} P_{j,T_z}+
\sum_{i\leq j,k\leq l} V^{T=0}_{J=1}(ij,kl) 
\sum_{J_z}D^+_{ij,J_z} D_{kl,J_z}.
\end{equation}
The pairing interactions are written in this case in terms of the non-collective pair operators (8,9), expressed in $jj$ coupling.
These interactions are not limited to the pairs with total $L=0$, as in the Hamiltonian (1),  and their matrix
elements are state-dependent. 
In Ref. \cite{qm_t1t0} it was shown that the ground state of the Hamiltonian (17)  can be described with high precision 
as a product of distinct quartets
\begin{equation}
|QM \rangle =\prod^{n_q}_{\nu =1}Q^{(\nu)\dag} |0\rangle \equiv \prod^{n_q}_{\nu =1} (Q^{(\nu)\dag}_1+Q^{(\nu)\dag}_0) |0\rangle .
\end{equation}
 The  collective isovector and isoscalar quartets introduced above have the same form as in Eqs.   (6,7) but with mixing
 amplitudes $x^{(\nu)}_{j_1 j_2}$ and $y^{(\nu)}_{j_1 j_2 j_3 j_4}$ which vary with the quartets.
Since in  (18)  the quartets are allowed to be different from each other, the  quartet model (QM) state (18) 
is expected to be a better approximation than the QCM state (4).  A comparison between the two
approximations will be presented in the next section.

 \section{Results}
 
 In the SU(4) limit and for degenerate single-particle orbits the QCM state (4) provides the exact solution
of the  Hamiltonian (1). In the following we shall examine how accurate the QCM ansatz remains beyond the
SU(4) limit and how  the spin-orbit interaction affects the competition between the isovector and isoscalar 
pairing correlations. In order to address these issues we first apply the QCM formalism to  the simple case 
of one orbit with  angular momentum $l$. More precisely,  we consider the orbit with  $l=3$ and we do calculations
for systems with 2, 4 and 6 proton-neutron pairs. To illustrate  the influence of the spin-orbit interaction on pairing 
correlations, we  perform calculations separately for the isovector and isoscalar  pairing Hamiltonians (assuming $g_1$=-1, $g_0$=0 and $g_1$=0, $g_0$=-1, respectively, in Eq. (1)) and we change progressively the energy splitting between the states $f_{7/2}$ and $f_{5/2}$ from 0 to $7$ MeV, which is about the spin-orbit energy
splitting used in realistic calculations in the $pf$ shell.  

In Fig. 1 we show how the pairing correlations energy, defined as the difference between the ground state energies calculated
without and with the pairing force, evolves with the spin-orbit energy splitting for the two pairing Hamiltonians. 
 When the spin-orbit interaction is zero, the two pairing correlations energies are equal to each other; they are also equal to the exact values since in 
the absence of spin-orbit interaction the exact solution for the isovector (isoscalar) Hamiltonian is a QCM state formed  
by  isovector (isoscalar) quartets. When the spin orbit is switched on,  the pairing correlations energies decrease. One can notice, however, that
this decrease is much stronger for the isoscalar pairing and also that it becomes more pronounced with increasing the number of proton-neutron pairs. 
\begin{figure}[h]
\begin{center}
\includegraphics*[scale=0.3,angle=-90]{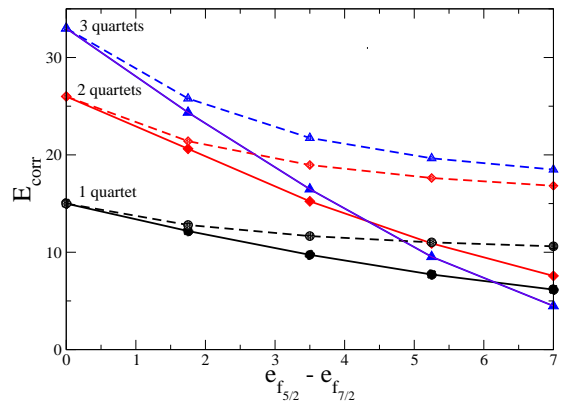}
\caption{Correlation energies (in MeV) provided by the QCM approach in corresponence with the Hamiltonian (1)  for 1, 2 and 3 quartets moving  in the orbits $f_{7/2}$ and $f_{5/2}$. Dashed lines refer to the isovector Hamiltonian ($g_1$=-1, $g_0$=0) while full lines refer to the isoscalar Hamiltonian ($g_1$=0, $g_0$=-1). On the horizontal axis we show the spin-orbit energy splitting between the two orbits (in MeV).}
\end{center}
\end{figure}

The suppression of the pairing correlations caused by the spin-orbit interaction can in fact be expected from the matrix elements 
of  the  $S=0$ and $S=1$ pairing forces calculated for the two-body wave functions  expressed in $jj$ coupling. 
Indeed, the diagonal matrix elements in these two channels for $j=l+1/2$ are given by
\beq
<(jj)J=0 | \sum_{T_z} \mathcal{P}^+_{T_z} \mathcal{P}_{T_z}| (jj)J=0>=\frac{2j+1}{2}
\eeq
\beq
<(jj)J=0 | \mathcal \sum_{S_z} \mathcal{D}^+_{S_z} \mathcal{D}_{S_z}| (jj)J=0>= \frac{(2j+1)(j+1)}{6j}
\eeq  
while, in the absence of spin-orbit coupling, both matrix elements are equal
to $(2l+1)=2j$. Thus, when the spin-orbit is so strong that the occupancy of 
 $j=l-1/2$ can be neglected, the matrix elements of the  $S=0$ and  $S=1$ pairing forces are 
 reduced, in the  limit of large $j$, by a factor 2 and 6, respectively.  The ratio of the
 matrix elements (19) and (20) is equal to $3j/(j+1)$, which is not very far from the ratio between the
 values of the two pairing energies shown in Fig. 1 at maximum spin-orbit energy splitting.
 
To probe the accuracy of QCM  results shown in Fig. 1, we have compared them with the exact results obtained by
diagonalisation. In all cases the relative errors for the  correlations energies, with respect to the exact values, turn out to be  small (below $1\%$),
except in the case of the isoscalar pairing energy at maximum spin-orbit coupling. For example, the errors for the system of 
6 proton-neutron pairs corresponding to the three values of spin-orbit energy splittings $\epsilon_{f_{5/2}}-\epsilon_{f_{7/2}}=\{3.50, 5.25, 7\}$ are 
equal, respectively, to $0.07\%$, $0.08\%$ and  $0.5\%$ for isovector pairing and to $0.04\%$, $0.7\%$ and $16.7\% $ for isoscalar pairing.
The large error for the isoscalar pairing at maximum spin-orbit energy splitting is related to the fact that  in this case the exact ground state 
of the isoscalar Hamiltonian is a state with $J=2$ rather than $J=0$.  Therefore the QCM result is compared not with the exact ground state of the 
isoscalar pairing Hamiltonian but with the first excited state  with $J=0$. A similar situation occurs in the limit of very large spin-orbit splitting 
when  all the nucleons occupy  the $f_{7/2}$ orbit. On the other hand, in this limit the QCM state provides an exact solution for the ground state of the
isovector pairing Hamiltonian.

Next we discuss the competition between the isovector and isoscalar  pairing correlations for different values of the two 
pairing strengths.  We assume a spin-orbit energy splitting equal to 7 MeV and parametrize the pairing strengths as $g_0=-(1+x)/2$ and $g_1=-(1-x)/2$, with $x$ ranging from -1 to 1.  We recall that in many
studies  the physical  value for the ratio between the two pairing strengths is assumed to be about 1.5
(e.g, see Ref. \cite{gerzelis_bertsch}), which corresponds to $x=0.2$. 
With the parameters given above we calculated the ground state energies provided
by the  QCM state (4) and by exact diagonalization. The comparison with the exact results shows that the accuracy of  QCM 
is decreasing with the increasing of the mixing parameter $x$. For example, the errors for the correlation energies 
corresponding to $x=\{-0.5, 0, 0.5\}$ are, respectively, 0.04$\%$, 0.1$\%$, and 2.1$\%$. This shows that  QCM 
provides accurate results for the physical region of the mixing parameter ($x \approx 0.2$).

The QCM results for pairing energies, defined by the average of the pairing forces in the
QCM state, are shown in Fig. 2a.  It can be observed that going from a  pure isovector force to  a pure isoscalar force
 the pairing energies evolve smoothly. The same behavior can be observed in Fig. 2b for the  
 overlaps  $<QCM|QCM(T=1)>$  and $<QCM|QCM(T=0)>$,  where the states $QCM(T=1)$ and $QCM(T=0)$ are the 
components of the $QCM$ state formed only with the isovector quartet (6) and, 
respectively, the  isoscalar quartet (7). The most important  message of Fig. 2  is that the isovector and isoscalar pairing 
correlations coexist for any  ratio of the two pairing forces. A similar  conclusion was obtained recently
for  the isovector and isoscalar pairing forces acting on time-reversed axially deformed states \cite{qcm_def}. It is worth
stressing that the majority of HFB calculations  for isovector-isoscalar pairing predict the coexistence of the two pairing
phases only for particular pairing forces and/or nuclei  \cite{goodman_prc,gerzelis_bertsch}.

\begin{figure*}[h]
\centering
\begin{tabular}{cc}
\epsfig{file=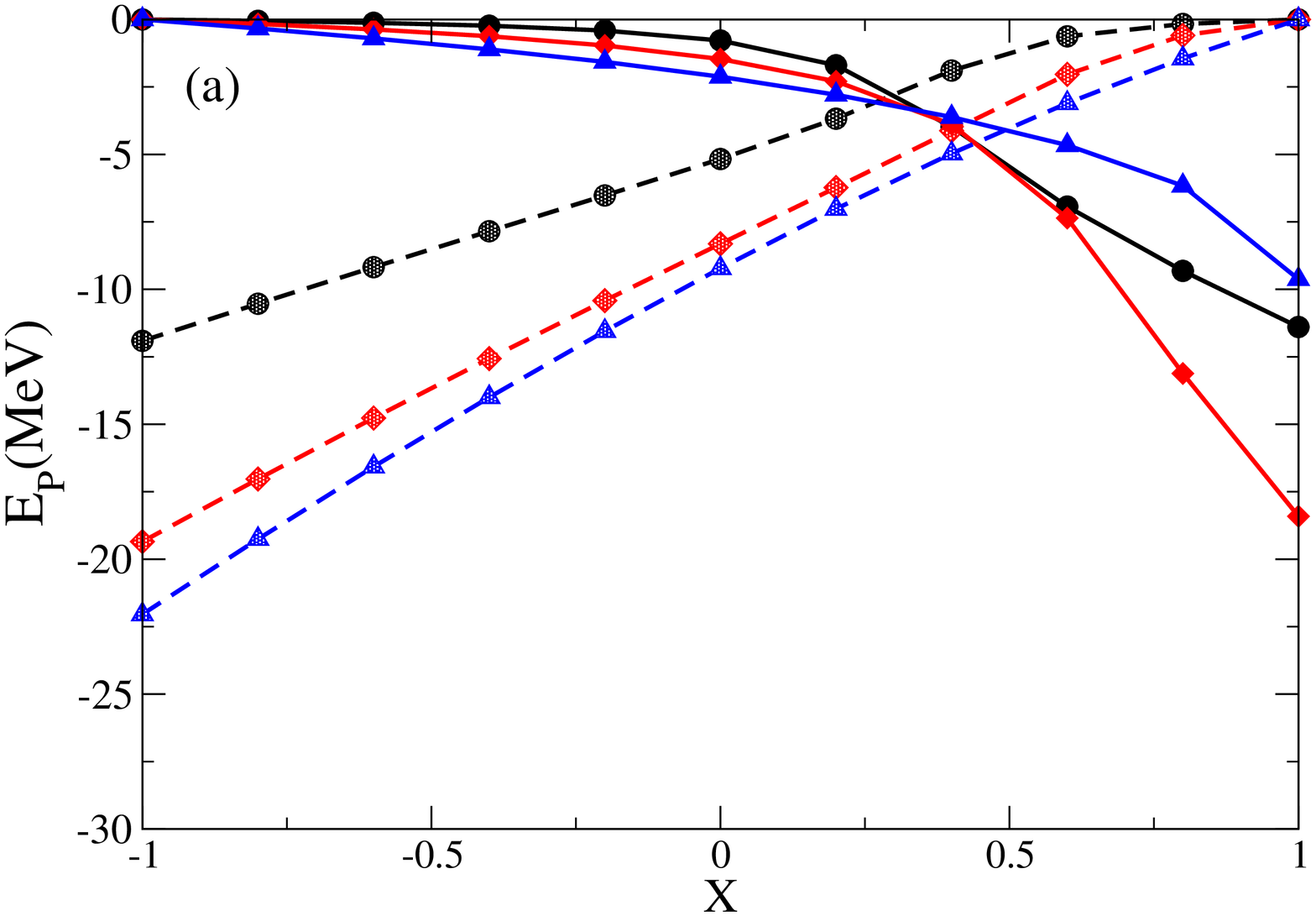,width=3in,height=3.55in,angle=-90} &
\epsfig{file=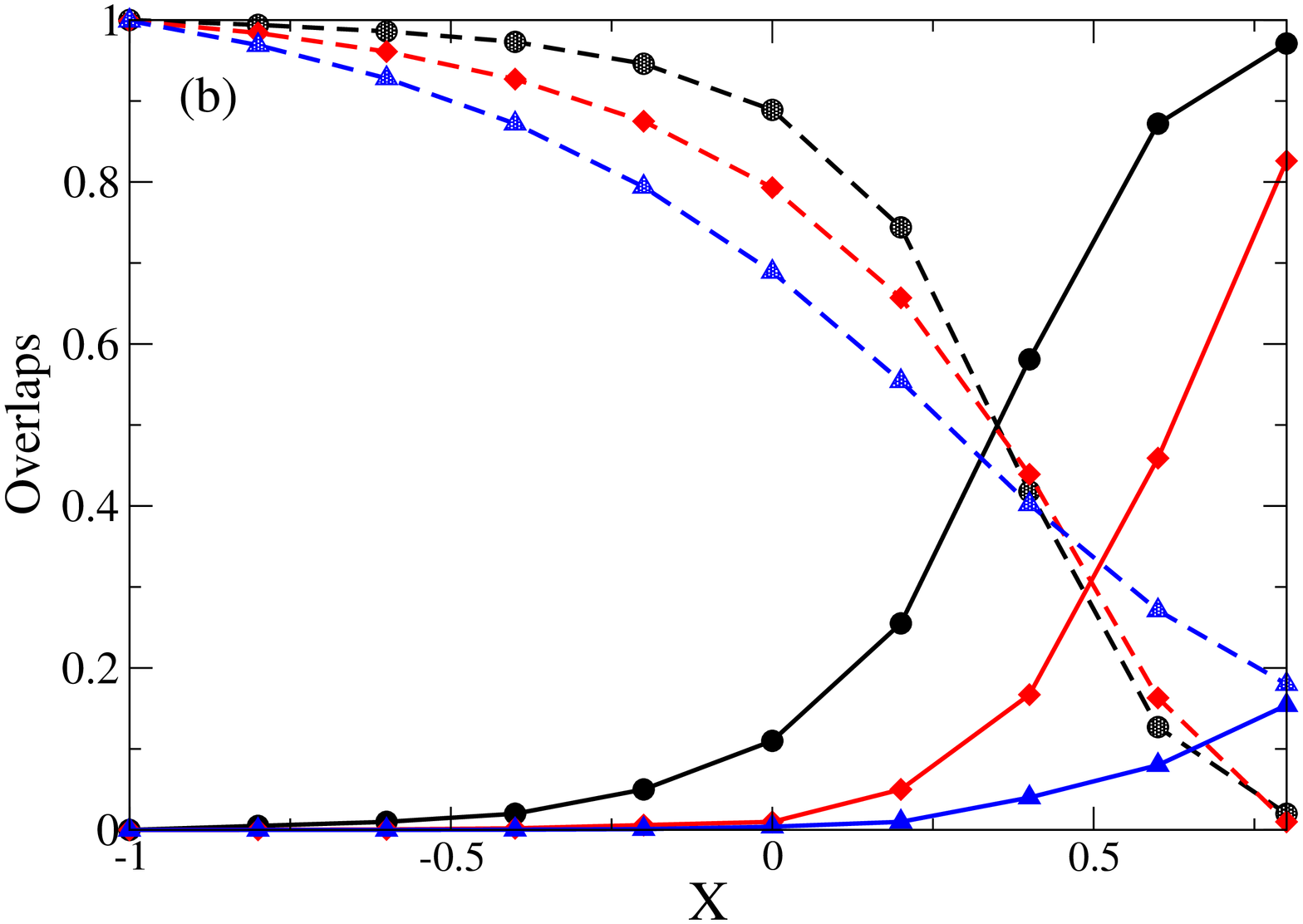,width=3in,height=3.55in,angle=-90} \\
\end{tabular}
\caption{ Pairing energies (left panel) and the overlaps  $<QCM|QCM(T=1)>$  and $<QCM|QCM(T=0)>$ (right panel)
as a function of the parameter $x$ which scales the strengths of the interactions (see text). The results refer to systems of
1, 2 and 3 quartets moving in the orbits $f_{7/2}$ and $f_{5/2}$ and they are shown with the same symbols adopted
in Fig. 1. The energy splitting between the orbits
is equal to 7 MeV.   }
\end{figure*}

In what follows we present applications of the QCM approach to  more realistic systems. As in previous
quartet model studies, we consider the $N=Z$ systems composed by nucleons moving in the valence shells 
above the cores $^{16}$O, $^{40}$Ca and $^{100}$Sn. First, we present the results obtained
with the pairing Hamiltonian (1). As strengths of the  pairing interactions we adopt the 
values $g_1=-0.377$ and $g_0=-0.587$ (in MeV). These pairing strengths have been suggested in
Ref.\cite{zuker} for the $pf$-shell nuclei. For the sake of simplicity, here we use the same pairing strengths for
all nuclei. In this way, keeping the ratio of the pairing strength unchanged, we can probe how the competition
between the isovector and isoscalar pairing correlations depends on the orbits and the spin-orbit splitting. 
The dependence of the pairing strengths on the atomic mass is taken into account through the factors $(18/A)^{0.3}$ for $sd$-shell nuclei,  
$(42/A)^{0.3}$ for $pf$-shell nuclei \cite{article_beta} and $(102/A)^{0.3}$ for nuclei above 
$^{100}$Sn.
For the three sets of nuclei the single-particle energies, which include the spin-orbit splitting, have been
taken as in shell model (SM) calculations performed with the interactions USDB  \cite{usd}, KB3G \cite{kb3g} and with the
G-matrix interaction of Ref. \cite{gmatrix}. The results are presented in Table I. In the second column we show
the pairing correlations energies and, in brackets, the errors relative to the exact results got by diagonalisation.
It can be seen that these errors are very small.  In Table I we also show the pairing energies
obtained by averaging the isovector and the isoscalar pairing forces in the QCM state. We notice that for all nuclei 
the pairing energies are significant in both channels. 

\begin{table}[hbt]
\caption{ Results of QCM calculations for $N=Z$ nuclei in various shells obtained  with the Hamiltonian (1) . 
$E_{corr}$ are the correlations energies while $E_P^{T}$ are the isovector ($T=1$) and isoscalar ($T=0$)
pairing energies obtained by averaging the pairing forces on the QCM state. In brackets we show the 
errors relative to the exact results obtained by diagonalisation. All energies are in MeV.}
 \begin{center}
\begin{tabular}{|c|c|c|c|}
\hline
\hline
 &  $E_{corr}$  & $E_{P}^{(T=1)}$ & $E_{P}^{(T=0)}$  \\ 
\hline
\hline
$^{20}$Ne &  4.005 (-)     &  -2.740  &  -2.390      \\ 
$^{24}$Mg &  5.914 (0.7\%)    &   -4.545  &   -2.660      \\ 
$^{28}$Si &   6.359 (0.5\%)        &  -4.389 &   -3.058        \\ 
\hline
\hline
$^{44}$Ti &  5.477  (-)        &  -3.486     & -4.478         \\ 
$^{48}$Cr &  8.571  (0.6\%)       & -6.946      & -4.985         \\ 
$^{52}$Fe &  9.812  (1.1\%)      & -8.576     &  -4.557         \\ 
\hline
\hline
$^{104}$Te &     6.413  (-)  &   -5.929    &  -2.229         \\ 
$^{108}$Xe &     11.195 (0.3\%) &  -10.860  &  -3.677           \\ 
$^{112}$Ba &     14.377 (0.5\%)  &  -14.376   &  -4.994          \\
\hline
\hline
\end{tabular}
\end{center}
\end{table}

Finally we have applied the QCM approach to the most general spherically symmetric pairing Hamiltonian (17). 
In Table II we present  the results of  QCM calculations performed by employing in this Hamiltonian
the same input as in Ref.\cite{qm_t1t0}. Namely, the single-particle energies and the pairing interactions  
for the three sets of nuclei shown in Table II are extracted from the shell model forces, respectively  USDB  \cite{usd}, 
KB3G \cite{kb3g} and G-matrix two-body force of Ref. \cite{gmatrix}.  In Table II the QCM results are compared 
to the exact  results and to the results of the QM approximation (18) presented in Ref. \cite{qm_t1t0}.
 One can see that the QCM approximation, in which it is supposed that all  quartets have
the same structure, gives accurate results, comparable with the QM approximation.
In Table II we present also the results obtained with the quartet condensate state 
$| \overline{QCM} \rangle = (\bar{Q}^+_1+\bar{Q}^+_0)^{n_q} |0>$ constructed with the 
quartets  introduced in Eqs.(10,11), expressed in terms of collective pairs.  One can notice that $\overline{QCM}$ 
describes less well the pairing correlations energies as compared to  QCM. On the other hand, 
as shown in  Ref. \cite{qm_t1},  the QCM and $\overline{QCM}$ approximations give very similar results when one
considers only the isovector interaction. 
That means that the isoscalar pairing force induces genuine four-body correlations which cannot be described
 accurately by a product of collective pairs. In Table II we also present the results given by the QCM states  
 constructed only by the isovector or the isoscalar quartets (10) and (11), i.e., 
 $| \overline{QCM}_{T=1} \rangle = (\bar{Q}^+_1)^{n_q} |0>$ 
 and $|\overline{QCM}_{T=0} \rangle= (\bar{Q}^+_0)^{n_q} |0>$.  It can be seen that, for the present pairing 
 interactions, the isoscalar pairing correlations are stronger than the isovector ones, with the exception of nuclei in the $sd$ shell. In all cases, however, as noticed also 
in the examples presented above, the isovector and isoscalar pairing correlations always coexist.

\begin{table}
\caption{  Results of various  calculations for $N=Z$ nuclei described by the Hamiltonian (17) . 
We show the correlations energies associated with the QCM state (4) and with the QCM approximations
defined with the quartets (10-11), i.e., $|\overline{QCM} \rangle = (\bar{Q}^+_1+\bar{Q}^+_0)^{n_q} |0>$,
 $|\overline{QCM}_{T=1} \rangle = (\bar{Q}^+_1)^{n_q} |0>$ and  $| \overline{QCM}_{T=0} \rangle= (\bar{Q}^+_0)^{n_q} |0>$.
 By $QM$ are indicated the results of Ref. \cite{qm_t1t0} obtained with the state (18).  
 In brackets we show the relative errors with respect to the exact results provided by diagonalisation.  
 All energies are in MeV.}
\begin{center}
\begin{tabular}{|c|c|c|c|c|c|c|}
\hline
\hline
 &  Exact & $QM$ & $QCM$  & $ \overline{QCM}$ & $\overline{QCM}_{T=1}$ & $\overline{QCM}_{T=0}$  \\
\hline
\hline
$^{20}$Ne   &   15.985  &   15.985 (  -   )             &     15.985 (  -   )                &     15.510 (2.97\%)   &     14.373  (10.08\%)   &    14.930    (6.60\%)     \\        
$^{24}$Mg  &   28.694   &    28.626 (0.24\%)   &     28.595  (0.34\%)   &      27.764   (3.24\%    &      23.229 (19.04\%)   &    26.299   (8.35\% )  \\        
$^{28}$Si    &   35.600   &    35.396 (0.57\%)  &     35.288 (0.88\%)    &       33.913 (4.74\%)   &      28.830 (19.02\% )  &    32.067    (9.92\%)  \\          
\hline
\hline
$^{44}$Ti    &    7.019   &     7.019  (   -   )              &   7.019 (   -   )                    &    6.302   (10.21\%)  &        6.273   (10.63\%)    &     4.825  (31.26\%) \\        
$^{48}$Cr    &  11.649  &   11.624 (0.21\%)   &   11.614 (0.30\%)          &   10.674    (8.37\%)   &       10.582  (10.67\%)   &     7.075    (39.26\%) \\        
$^{52}$Fe    &  13.887  & 13.828   (0.42\%)  &  13.799  (0.63\%)          &  12.971      (6.60\%)   &    12.795 (7.92\%)    &      9.589   (30.95\%) \\        
\hline
\hline
$^{104}$Te   &    3.147   &  3.147 (   -   )                &    3.147  (   -   )                  &   3.052  (3.02\%)       &   3.041 (3.37\%)    &   1.512   (51.95\%) \\        
$^{108}$Xe   &   5.505   &   5.495 (0.20\%)   &    5.489 (0.29\%)         &  5.279   (4.10\%)       &   5.239 (4.83\%)    &   2.530 (54.04\%) \\        
$^{112}$Ba    &  7.059    & 7.035   (0.34\%)    &    7.017  (0.59\%)         &   6.691  (5.21\%)       &   6.609 (6.37\%)   &   4.391   (37.79\%) \\       
\hline
\hline
\end{tabular}
\end{center}
\end{table}

The importance of the mixing between various  pairs to preserve exactly the isospin and
the angular momentum of the ground state can be  seen by comparing the predictions of  QCM 
with the PBCS approximations (14-16).  As seen in Table III, the  results of the PBCS approximations
(14-15) and (16), which do not conserve the isospin and the angular momentum, respectively,  are much
less accurate than the ones provided by QCM. We also notice that for all nuclei $PBCS1$ gives more binding than $PBCS_{iv}$,
while the latter gives more binding than $PBCS_{is}$, except for $sd$-shell nuclei. Thus, for  $sd$-shell
nuclei the condensate of isoscalar proton-neutron pairs appears to be favorite with respect to the condensate of isovector pairs while the opposite is true in the other shells.
On the other hand,
from Table II one also sees that the approximation $\overline{QCM}_{T=1}$, in which the isovector pairs are
mixed together to form a state with good isospin, gives in $sd$-shell nuclei more binding than $PBCS_{is}$.
It is therefore clear that $PBCS_{is}$ results are not by themselves enough to conclude
that the ground state of $sd$-shell nuclei is mainly
described by a condensate of isoscalar proton-neutron pairs. In fact, as shown in Table II and Table III, a proper
description of the competition between the isovector and isoscalar pairing correlations requires a ground
state in which all types of pairs are mixed together in order to conserve  exactly the spin and the angular momentum. 

\begin{table}
\caption{ Correlations energies provided by the PBCS-type states (14-16) in comparison with the QCM  results. 
 In brackets we show the relative errors with respect to the exact results obtained by diagonalisation.
 These results have been obtained with the  Hamiltonian (17) using the interactions described in the text.  All energies are in MeV.}
\begin{center}
\begin{tabular}{|c|c|c|c|c|}
\hline
\hline
 & $QCM$  & $PBC1$ & $PBCS0_{iv}$ & $PBCS0_{is} $ \\
\hline
\hline
$^{20}$Ne     &     15.985 ( - )     &   14.011 (12.35\%)        &    13.664   (14.52\%)         &     13.909  (12.99\%) \\        
$^{24}$Mg    &     28.595  (0.24\%)   &    21.993 (23.35\%) &   20.516   (28.50\%)  &    23.179     (19.22\%)       \\        
$^{28}$Si     &     35.288    (0.57\%)  &     27.206 (23.58\%)  &   25.293   (28.95\%)   &    27.740  (22.19\%)       \\          
\hline
\hline
$^{44}$Ti    &   7.019 ( - )   &   5.712  (18.62\%)    &   5.036   (28.25\%)   &    4.196     (40.22\%)         \\        
$^{48}$Cr   &   11.614  (0.21\%)  &    9.686  (16.85\%)   &     8.624   (25.97\%)   &     6.196   (46.81\%)       \\        
$^{52}$Fe   &  13.799   (0.42\%)  &     11.774   (15.21\%) &     10.591  (23.73\%)  &       6.673  (51.95\%)        \\        
\hline
\hline
$^{104}$Te    &    3.147 ( - ) &     2.814  (10.58\%)    &     2.544  (19.16\%)   &     1.473   (53.19\%)    \\        
$^{108}$Xe   &  5.489  (0.20\%)  &      4.866   (11.61\%)  &     4.432   (19.49\%)   &    2.432    (55.82\%)  \\        
$^{112}$Ba    &  7.017  (0.34\%)   &    6.154   (12.82\%)   &    5.635   (20.17\%)   &    3.026   (57.13\%)  \\       
\hline
\hline
\end{tabular}
\end{center}
\end{table}

\section{Summary and Conclusions}
In this paper we have generalized the quartet condensation model for the treatment of spherically symmetric 
isovector $(T=1,J=0)$ and isoscalar $(T=0,J=1)$ pairing forces. The basic assumption of the QCM approximation
is  that the ground state correlations induced by these forces can be described in terms of products of identical quartets formed by coupling two neutrons and two
protons to total isospin $T=0$ and  total angular momentum $J=0$. The generalized QCM approach has been first applied to pairing forces formulated in terms of 
isovector $(T=1,S=0,L=0)$  and isoscalar $(T=0,S=1,L=0)$ pairs. For these
forces we have illustrated how the  spin-orbit interaction affects the pairing correlations and we have studied the competition between
the isovector and isoscalar pairing. Then, the QCM approach has been applied to realistic systems described by the most general pairing Hamiltonian formulated in terms of $(T=1,J=0)$ and $(T=0,J=1)$ pairs. We have shown that for both Hamiltonians the QCM gives an accurate
description of the pairing correlations. We have also shown that in the QCM approximation the correlations in the two pairing 
channels coexist for any admixture of isovector and isoscalar pairing forces, which confirms the findings of Refs. \cite{qm_t1t0,qcm_def}.

We wish to conclude this paper by emphasizing the striking analogy between the like-particle and 
proton-neutron pairing pictures which has emerged in this study and which is also supported by our
previous works on the same subject \cite{qcm_t1,qcm_ngz_t1,qcm_wigner, qm_t1,qm_t1t0,qcm_def}. 
Thus, if on one side a condensate of collective  $J=0$ pairs provides a good approximation to 
the ground state of spherically  symmetric like-particle pairing Hamiltonians, on the other side, 
as shown here, a condensate of $J=0$,$T=0$ quartets provides a good approximation to  the
ground state of spherically symmetric proton-neutron pairing Hamiltonians. In the
case of proton-neutron pairing, then, collective quartets appear to play the same
role as Cooper pairs in the case of like-particle pairing. A basic difference between the like-particle
pairing and pairing in $N=Z$ systems is that in the latter one needs to couple the isospin and the spin of the
pairs in order to construct wave functions with well-defined total isospin and total angular momentum. 
As demonstrated in this paper, in even-even $N=Z$ nuclei the  quartets built by coupling 
two pairs to $T=0$ and $J=0$ do represent the simplest form of many-body structures whose condensate 
can guarantee a ground state with total $T=0$ and total  $J=0$. The fact that in the quartet condensate state, 
which describes accurately the pairing forces in $N=Z$ nuclei,  the isovector and isoscalar proton-neutron pairing correlations are 
strongly entangled indicates that  it might be difficult to disentangle them by proton-neutron transfer reactions. 
If in open shell $N=Z$ nuclei the quartets are indeed strongly correlated structures acting coherently as a condensate, 
one would expect collective features for alpha-particle transfer  reactions (e.g., significant enhancement of the transfer 
with the number of quartets) rather than for the transfer  of  proton-neutron pairs.

\vskip 0.4cm
\noindent
{\bf Acknowledgements}
\vskip 0.2cm
\noindent
We thank D. Gambacurta for having provided us with the shell model results discussed in the text.
This work was supported by the Romanian National Authority for Scientific Research,
 CNCS UEFISCDI,  Project Number PN-II-ID-PCE-2011-3-0596.

\end{document}